\documentclass[preprint,prl,showpacs,preprintnumbers,superscriptaddress,amsmath,amssymb,floatfix]{revtex4-1}  

\usepackage{amsmath,dcolumn,graphicx,epsf,ulem} 
\usepackage{natbib}
\setkeys{Gin}{width=8.5cm,keepaspectratio}   
\usepackage{dcolumn}
\usepackage{bm}
\usepackage{xcolor}
\usepackage{siunitx}

\begin{document}   
   
\title{Geometric influence on the net magnetic moment in LaCoO$_3$ thin films}   
  
\author{T. Joshi}
\affiliation{Department of Physics, University of California, Santa Cruz, CA, 95064, USA}
\author{D. P. Belanger}
\affiliation{Department of Physics, University of California, Santa Cruz, CA, 95064, USA}
\author{Y. T. Tan}
\affiliation{Sibley School of Mechanical and Aerospace Engineering, Cornell University, Ithaca, NY, 14853, USA}
\author{W. Wen}
\affiliation{Shanghai Institute of Applied Physics, Chinese Academy of Sciences, Pudong New District, Shanghai, 201204, China}
\author{D. Lederman}
\affiliation{Department of Physics, University of California, Santa Cruz, CA, 95064, USA}

\date{\today}   
   
\begin{abstract}
The different magnetic behaviors of LaCoO$_3$ films grown on LaAlO$_3$ and SrTiO$_3$ are related to the Co-O-Co bond angles and the constraints imposed on the Co-O bond lengths by the substrate geometries.  Long-range magnetic order occurs below $T \approx 90$~K when the Co-O-Co bond angle is greater than 163$^\circ$, consistent with the behavior of bulk and nanoparticles forms of LaCoO$_3$.  A LaAlO$_3$ substrate prevents magnetic long-range order at low temperatures near the film substrate interface and collinear antiferromagnetic sublattices away from the interface.  At low temperatures, the antiferromagnetically ordered sublattices are non-collinear in films grown on SrTiO$_3$ substrates, leading to a significant net moment.

\end{abstract}    
   
   
\maketitle   
   
\section{Introduction}

LaCoO$_3$ (LCO) in bulk crystalline form exhibits a small net moment below a temperature of $T\approx 89$~K. The moment arises from an asymmetry of the the antiferromagnetic sublattices induced by the structural geometry at twin plane interfaces along (100) pseudocubic planes of the lattice \cite{kbyfwmy18}.  Away from the twin boundaries, for temperatures above 100~K, fits to Curie-Weiss behavior indicate antiferromagnetic interactions \cite{dbbycfb13}, with $\theta_{CW} \approx -182(4)$~K, between atomic moments of 3.45(2) $\mu_B$, yet bulk LCO does not order antiferromagnetically at low temperatures because the interactions are reduced before the transition can take place.  The strength of the magnetic interactions has been correlated with the angle of tilt between adjacent oxygen octahedra surrounding Co ions, given by the Co-O-Co bond angle.  As the temperature decreases towards $T=40$~K, the magnetic interaction weakens as the Co-O-Co angle decreases towards a critical angle of 163$^\circ$ \cite{bkbdmzmb16,dbhycfbab15,dhbcyfab15_2,dbbycfb13}.  The critical angle of 163$^\circ$ found experimentally is consistent with results from generalized gradient approximation calculations \cite{lh13}.  When the temperature is below $T=40$~K, the angle no longer changes and the interaction strength is insufficient to support ordering.  The magnetic state is thermally activated above $T=40$~K \cite{lh13,bkbdmzmb16}.  This mechanism does not rely on proposed \cite{kesaks96} thermally excited intermediate spin state ($ t^{5}_{2g}e_{g}^{1} $) transitions that could become possible with Jahn-Teller-like distortions of Co-O bonds in the oxygen octahedra; neutron scattering and EXAFS studies found no significant evidence of the necessary distortion of the octahedra in bulk or nanoparticle forms of LCO \cite{sjabbbmpz09,jbsbamz09}. 

The correlation between the behavior of antiferromagnetism and the Co-O-Co angle applies consistently for bulk and nanoparticle forms of LCO \cite{kbyfwmy18,bkbdmzmb16,dbhycfbab15,dhbcyfab15_2,dbbycfb13}, but the magnetization of thin film samples has not previously been examined in that context.  Here we show evidence that this correlation does hold in films, even though the film geometries are distorted by the strain from the substrates upon which they are grown.  This gives a common basis for understanding why strikingly different magnetic behaviors appear under different strain conditions.

A large number of studies have addressed the magnetic behavior of LaCoO$_3$ films on SrTiO$_3$ (STO) and LaAlO$_3$ (LAO) substrates   \cite{zwjzklx20,fmzlchwgmzll19,gdkhwfl19,lwzrsjcwwghggjg19,snmskwpkr18,zzylhwskyss16,mbjavs15,qjsgxmvtfzlpzlbbb15,fydnkmnkat15,kckjzlk14,ckjhrmshshkl12,srkkmsw12,rhsd10,mlwcas09,fapssl08,hrsd09,prkkmsfw09,fms08}.  A common starting point for many studies of LCO films on STO substrates, in which a net spontaneous moment occurs, has been with the assumption that the interactions in the film are ferromagnetic.  Likewise, it has often been stated that LCO films on LAO substrates are paramagnetic.  A model has not been developed that adequately explains why ferromagnetic interactions would appear in LCO films on STO substrates, but not on LAO substrates and not in bulk or nanoparticle forms of LCO.  One of the striking characteristics of LCO films, as well as bulk and nanoparticle forms of LCO, is that long-range order, when it occurs, tends to occur within temperature range $65<T<90$~K.  This suggests a common interaction, regardless of whether a net moment appears, and the interactions in the films are likely antiferromagnetic as they are in other forms of LCO.  We will show, using magnetometry and structural data, that the films behave in ways consistent with bulk LCO in that the interactions are antiferromagnetic, but only when the Co-O-Co bond angles are greater than approximately 163$^\circ$.  The substrates strain the LCO films on STO substrates in a way that induces a net moment via antiferromagnetic sublattices that are non-collinear.  In contrast, close to the interfaces of LCO films on LAO substrates, magnetic interactions for bonds in planes parallel to the interfaces are suppressed and no magnetic long-range order takes place at low temperatures.  Further from the LCO/LAO interface, the LCO lattice relaxes towards the bulk LCO structure and this allows ordering with antiferromagnetic collinear sublattices, resulting in no net moment.  In the geometric interpretation, it is the substrate strain that controls the appearance of the net moment in LCO films through non-collinear alignment of antiferromagnetic sublattices, not the appearance of ferromagnetic interactions.  Switching between a state with no net magnetic moment to one that has a significant net moment, as has been observed in devices~ \cite{hppjdy13}, is controlled by the nature and size of the strain imposed by the substrate.  We will present this geometric interpretation correlating the Co-O-Co angle to the appearance of a net moment using data obtained from x-ray scattering and magnetometry experiments on thin LCO films.

\section{Experimental Details}

Before growth, single crystalline SrTiO$_3$ (STO) (001) and LaAlO$_3$ (LAO) (001) substrates were cleaned by sonicating in acetone and isopropanol for 10 min each.  In addition, STO substrates were annealed twice at 1000~$^\circ$C for 2~hrs in an atmospheric annealing furnace (Lindberg/Blue M), followed by a 30~sec deionized water etch as discussed in Ref.~\cite{CIESS2012}. The procedure results in a TiO$_2$-terminated atomically flat surface with a step-terrace structure. 

All films were grown using a pulsed laser deposition (PLD) technique.  A stoichiometric LaCoO$_3$ target was ablated using a 248-nm wavelength KrF excimer laser (Coherent Compex Pro F 102).  Samples were mounted to the substrate holder using high vacuum compatible conductive silver paint (Ted Pella, Inc.) to maximize thermal contact with the substrate heater. During growth, the substrate temperature was kept at 650~$^\circ$C, and the chamber pressure was maintained at 200~mTorr with 120~sccm constant flow of ultrahigh purity  O$_2$ gas.  After growth, the sample temperature was decreased to room temperature at a rate of 20~$^\circ$C/min in an ambient growth pressure.

Crystal quality was monitored during sample growth using the \textit{in-situ} reflection high energy electron diffraction (RHEED) technique and sample topography was measured after growth using tapping mode atomic force microscopy at room temperature (Oxford Cypher AFM).  Epitaxial strain on the films was quantified using x-ray diffraction spectra measured using a Rigaku Smartlab using Cu K$_{\alpha_1}$ radiation.  Thin films lattice parameters were calculated using Bragg peaks fitted to a Gaussian function and the film thickness was estimated using low-angle x-ray reflectivity data analyzed using GenX software \cite{Bjorck:aj5091}.  In addition, X-ray scattering data were obtained at 10 keV energy at BL14B1 of the Shanghai Synchrotron Radiation Facility (SSRF) \cite{Wen2015}. Two LCO films grown on LAO substrates and two grown on STO substrates were measured using Huber 5021 six-circle diffractometer \cite{Gao:kc5032}.    

Magnetometry data were taken using a superconducting quantum interference device (SQUID) magnetometer (Quantum Design MPMS XL) with the reciprocating sample option (RSO) method of measurement.  Each sample was mounted with the field oriented in the plane as well as out of the plane.  The sample was cooled in a $\mu_0 H=100$~mT field from 320~K to 5~K at a rate of 5~K/min and the magnetization was measured at a temperature interval of 1~K while cooling the sample at a rate of 2 K/min.  To eliminate the diamagnetic contribution from the substrates, after each measurement of a LCO film on the substrate, the film was etched off the substrate by chemically dipping the sample into aqua-regia for 30 seconds. The magnetic measurements were then repeated on the substrate with both field orientations with identical measurement conditions.  After proper subtraction of the signal from the substrate, the data were fitted to the Curie-Weiss model with an additional constant background term. 

\section{Results and Discussion}
\subsection{Structural Characterization}

\begin{figure}
    \centering
  
  \includegraphics{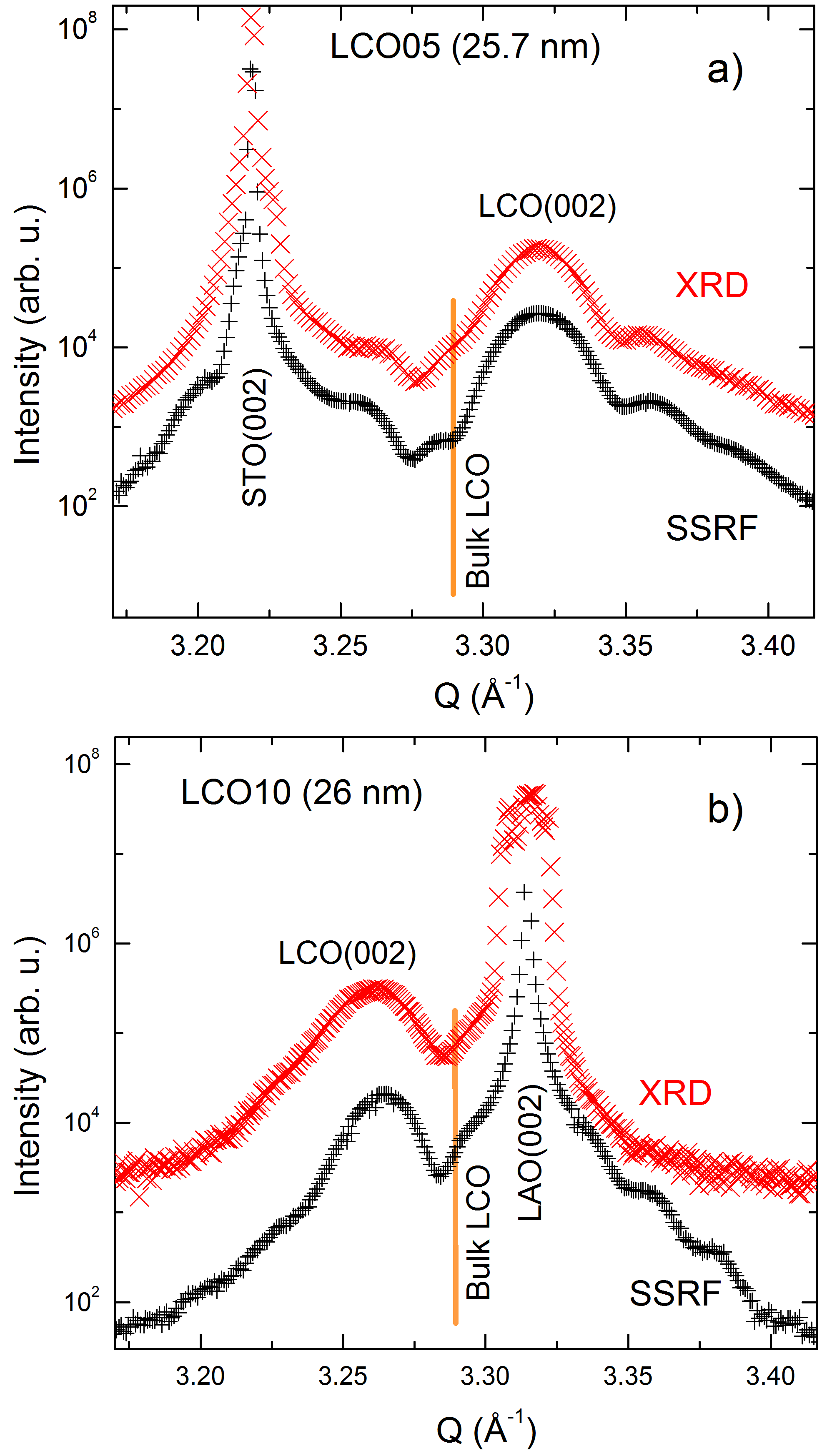}
  
    \caption{X-ray diffraction (XRD) and high-resolution synchrotron spectra (SSRF) for a 25.7~nm thick LCO film on STO (001) (a) and a 26~nm thick film on LAO (001) (b).The orange solid lines represent the bulk peak position.
    \label{fig:XRD1}}
\end{figure}

\begin{figure}
    \centering
  
  \includegraphics{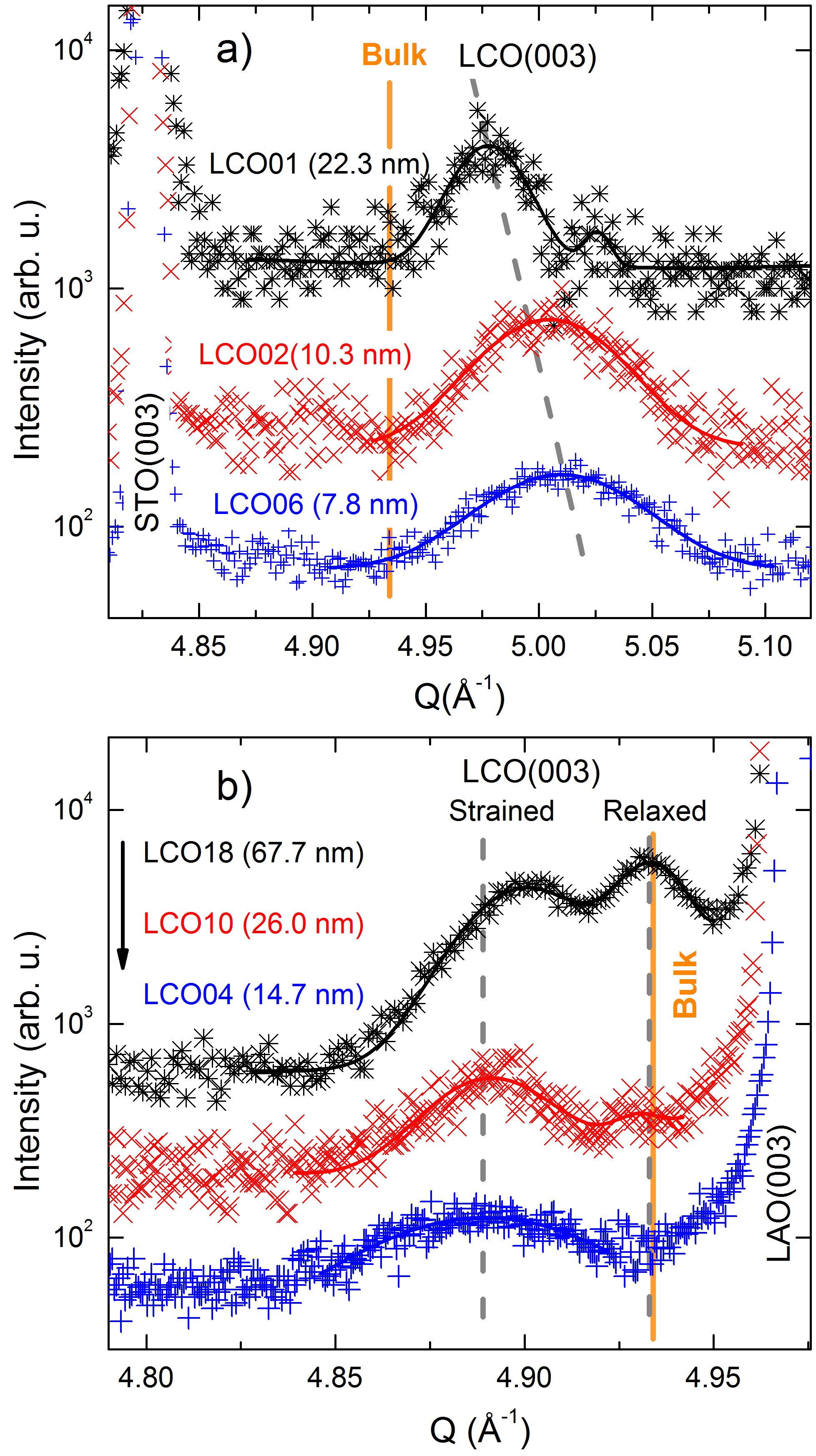}
  
    \caption{(a) X-ray diffraction spectra measured for LCO films on STO substrates. The gray dashed line is a visual guide for the shift in the film peak position with increasing film thickness. The shift corresponds to the different strain states in the film. (b) X-ray diffraction spectra measured from LCO films grown on LAO substrates. The two dashed lines indicate the peak positions of the strained (left) and relaxed part of the film in thicker films. The solid curves are Gaussian fits to the LCO peaks. 
    \label{fig:XRD2}}
\end{figure}

X-ray diffraction and synchrotron spectra for a 25.7~nm thick LCO film on STO (LCO05) and a 26~nm thick film on LAO (LCO10) are shown in Fig.~\ref{fig:XRD1}(a) and (b), respectively.  The synchrotron data are higher resolution and show interference peaks clearly.  The shift in the Bragg peak position of the films relative to bulk LCO (vertical solid line) corresponds to a change in $c$ lattice parameter originating from the tensile strain on the LCO film parallel to the substrate induced by STO and compression parallel to the substrate induced by LAO.  Three of the films with different film thicknesses grown on STO substrates are shown in Fig.~\ref{fig:XRD2}(a) and Fig. S1 (Supplemental Material).  The spectra show a progressive shift in the LCO~(003) peak position toward the left side of the figure due to the partial lattice relaxation with increased film thickness.  The LCO05 sample (25.7~nm film) shows a split in the peak corresponding to the relaxed and strained portions of the film. The reciprocal space map of (103) reflection of LCO05 film is shown in Fig. S2 (Supplemental Material).  Also, AFM images of LCO films on STO are listed in Fig. S3 (Supplemental Material).  Films grown on LAO, which have a relatively smaller film-substrate lattice mismatch, on the other hand, showed a smaller peak shift with increased film thickness [Fig.~\ref{fig:XRD2}(b)].  The peak splitting for the thicker film was observed due to the lattice-relaxation, similarly to the film grown on STO.  

The thickness of the films calculated from x-ray reflectivity (Fig. S4, supplemental Material) and synchrotron spectra are listed in table \ref{table:film_parameters}.  The thickness values calculated from synchrotron data are consistently smaller than those calculated from x-ray reflectivity because the x-ray diffraction data are sensitive to the crystalline coherence length, whereas the reflectivity is not sensitive to the crystallinity, but rather to surface and interface roughness.

The room temperature $c$ and $a$ lattice parameters of all of the LCO films on LAO and STO substrates are also listed in Table \ref{table:film_parameters}.  The $c$ lattice parameters are calculated using x-ray Bragg diffraction peaks. The values for the $a$ lattice parameters shown are well-known values of the substrates except the relaxed phase represented by the second peak in LCO18; the film in-plane lattice parameters might differ from the substrate in-plane lattice parameters in that case due to the lattice relaxation.

\begin{table}

        \caption{LCO films on LAO and STO substrates including film thicknesses determined from x-ray reflectivity (t$_{XRR}$) and synchrotron x-ray diffraction (t$_{SSRF}$) and room temperature $c$ and $a$ lattice parameters. The values of $a$ correspond to the in-plane lattice constants of the the substrate. Uncertainties of the last digit are in parentheses.}
        \vspace{5mm}
        
\begin{tabular}{|c|c|c|c|c|c|}

\hline

	Sample (Substrate)            &  t$_{XRR}$ (nm) & t$_{SSRF}$ (nm) & $c$ (\AA) & $a$ (\AA) & $c/a$\\ \hline
	LCO08 (LAO)     &11.5(2) & -- & 3.877(8) & 3.791&1.023(2)\\
	LCO04  (LAO)    &14.7(3) & 10.4(28) &3.865(6) & 3.791&1.022(2)\\
	LCO10  (LAO)    &26.0(2) & -- &3.868(7) & 3.791&1.020(2)\\
	LCO18  (LAO)    &67.5(4) & 23.0(20)&3.846(4) & 3.791& 1.015(1)\\
	LCO18 (2nd peak)    &67.5(4) & --&3.818(5)& - & -\\
    LCO06 (STO) & 7.8(3) & 6.0(3)& 3.763(8) & 3.905 & 0.964(1)\\
    LCO02 (STO) & 10.3(5) & --&3.773(5) & 3.905 & 0.966(1)\\
    LCO01 (STO) & 22.3(3) & --&3.787(1) & 3.905 & 0.970(1)\\
    LCO05 (STO) & 25.7(2) & 19.1(16) &3.786(3) & 3.905 & 0.970(1) \\
    LCO20 (multilayer) & 40.7(10) & --& 3.789(9) &-&- \\

\hline

\end{tabular}
\label{table:film_parameters}
\end{table}

\subsection{Magnetization}

\begin{figure}
        \includegraphics[width=5in]{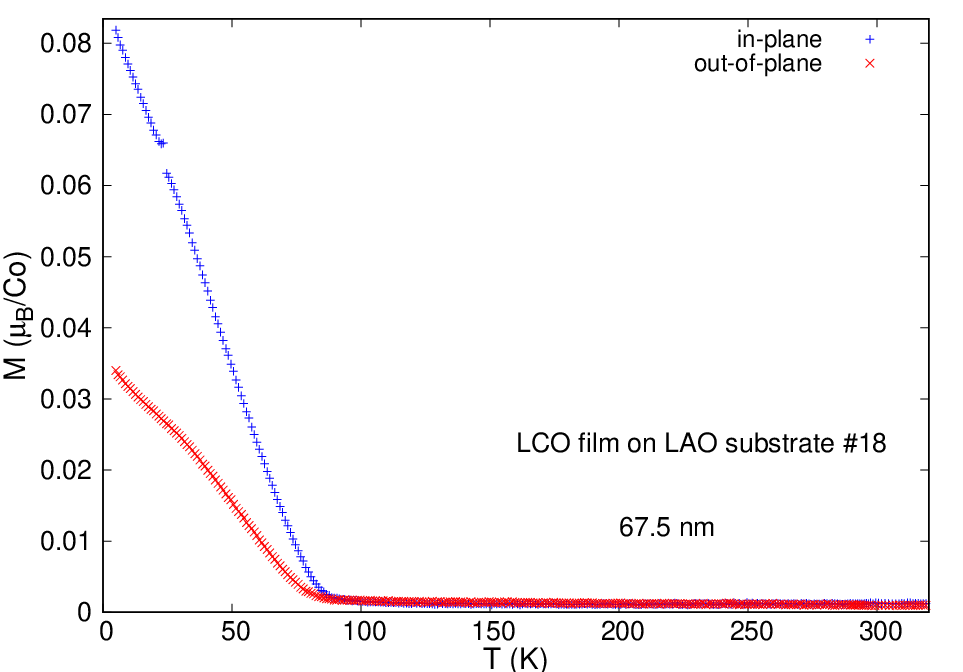}
        \caption
        {$M$ vs $T$ for the 67.5~nm (LCO18) film on a LAO substrate with $\mu_0 H=100$~mT. The signal from the substrate alone has been subtracted from that of the film and substrate.  Particular attention was paid to the stability of the measurements in this case to demonstrate that the signal for $T>100$~K is small with little temperature dependence; no fit was possible to the Curie-Weiss approximation in Eq.\ \ref{CW_fit}.
}
\label{fig:LCO18}
\end{figure}

\begin{figure}
        \includegraphics[width=5in]{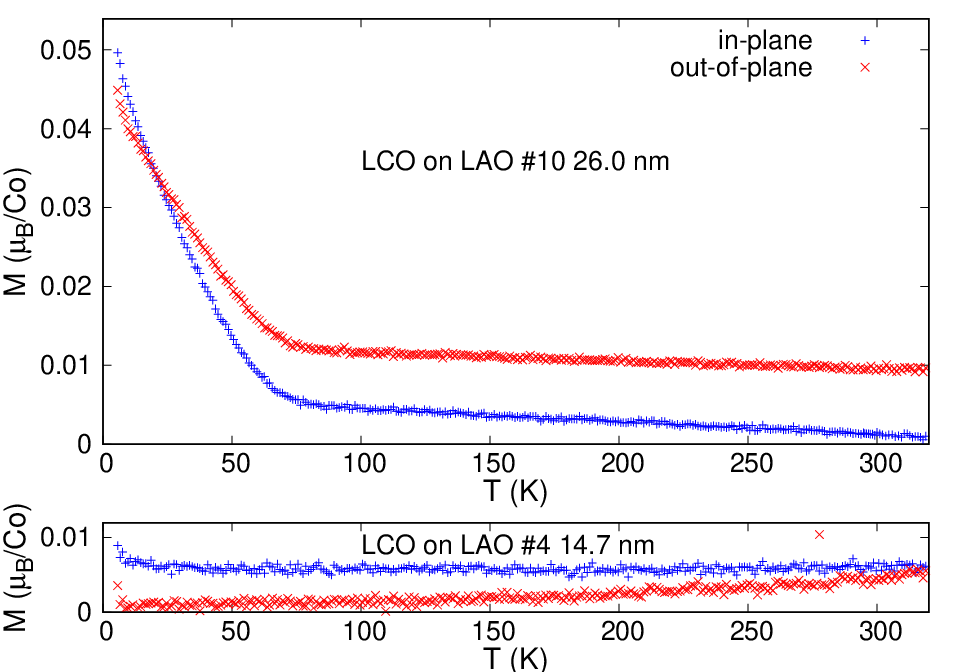}
        \caption
            {$M$ vs $T$ for the 26.0~nm (LCO10) (a) and 14.7~nm (LCO04) (b) films on LAO substrates with $\mu_0 H=100$~mT. The substrate signal has been subtracted as described in the text, however, the stability of the SQUID was not as good for these data sets, so the background subtraction shows small residual signals. Clearly, the magnetic ordering for $T<100$~K is smaller for the 26~nm film (a) relative to the 67.5~nm film shown in Fig.\ \ref{fig:LCO18}, and it is absent in the 14.7~nm film (b).
}
\label{fig:LCO10_LCO04}
\end{figure}

As a result of the small volume of the LCO film relative to that of the substrate, the significant diamagnetic contribution present in the magnetometry data of the combined film and substrate must be accounted for in the analysis of the LCO film magnetic behavior.  To accomplish this, after measurements on an LCO film on the substrate, the film was etched off chemically by dipping the sample into aqua-regia, as discussed in the experimental section, and the magnetic measurements were repeated on the substrate with both field orientations and identical measurement conditions.

To analyze the magnetic behavior of the films over the wide temperature range $5<T<320$~K, it works well to subtract the magnetic substrate signal by simply subtracting the substrate magnetic data directly.  To analyze the LCO film magnetism on STO substrates well above the transition, where the signal is relatively weak, we used the fits to the substrate data and subtracted them from the data taken with the films on the substrates.  For LCO01 and LCO05 samples, data taken under conditions of high thermal stability fit well to a Curie-like dependence plus a constant, given by 
\begin{equation}
	\frac{M}{H} = \frac{C}{\left(T-\Theta_{CW}\right)} +B \quad ,
	\label{CW_fit}
\end{equation}
over the range $90<T<320$~K, where $C$ is a constant, $\Theta_{CW}$ is the Curie-Weiss temperature, and $B$ is a background. Using these fits to analyze the film magnetism over this temperature range yielded essentially the same results as when the substrate were directly subtracted from the substrate data, but using the fits resulted in slightly less noise and a smaller residual background.  For LCO films on LAO, even with high thermal stability, we simply subtracted the substrate data directly over this range because we could not obtain fits to
a critical behavior as a function of $T$ for $T<T_N$ given by 
\begin{equation}
    M = A(T_N-T)^\beta + B\quad ,
    \label{crit-fit}
\end{equation}
where $T_N$ is the critical transition temperature and $B$ is a constant background, without adding at least one additional fitting parameter, and doing so did not aid the fits to the LCO film magnetization.

The magnetic behavior of the 67.5~nm (LCO10) film, after subtracting the signal from the LAO substrate, is shown in Fig.\ \ref{fig:LCO18}  for $5<T<320$~K for $\mu_0 H=100$~mT approximately in the plane of the substrate and perpendicular to it.  For this sample, particular attention was paid to the thermal stability while taking data.  This was done in an attempt to characterize the small signal for $T>100$~K using Eq.~\ref{CW_fit}, but no credible fits were possible. The magnetization is weak and the Curie-Weiss temperature is negative, so the signal has little variation with temperature over the fitting range; the Curie-Weiss temperature is highly correlated with the constant background in the fits, leading to very large parameter uncertainties. Below $T \approx 100$~K, a small moment appears and is about twice as large in the in-plane measurements.  The data do not follow critical behavior expected for an order parameter conjugate to the applied field close to the transition temperature $T_N$ given by Eq.~\ref{crit-fit}, 
but rather indicate a spin-flop-like signal associated with antiferromagnetic ordering below $T_N \approx 90$~K.  Such behavior is to be expected for antiferromagnetic order in small applied fields when there is small or negative uniaxial or cubic anisotropy~\cite{kbyfwmy18}.  In contrast to the bulk, however, the LCO film magnetic interactions extend to low temperatures.  We conclude that the moments form two sublattices that order collinearly.  The transition temperature is consistent with the size of the magnetic moments and the interaction strength found in bulk LCO at temperatures between $150$~K and room temperature and at all temperatures below room temperature in LCO nanoparticles \cite{kbyfwmy18,bkbdmzmb16,dbhycfbab15,dhbcyfab15_2,dbbycfb13}.

\begin{figure}
        \includegraphics[width=5in]{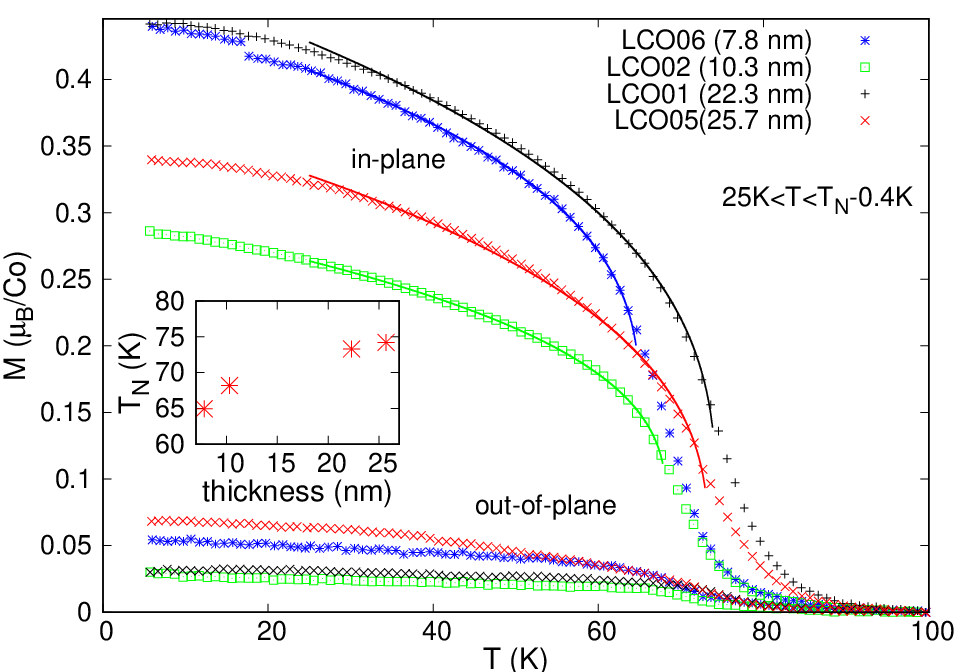}
        \caption{
        $M$ vs $T$ for four LCO films on STO substrates measured in a magnetic field of $\mu_0 H=100$~mT.  The upper data sets are for $H$ in the plane of the film and the lower sets are for $H$ perpendicular to the plane of the film. The substrate backgrounds are subtracted and the data are then adjusted vertically so that they pass through zero at $T=100$~K, so the curves approximately represent the contribution from long-range magnetic ordering.  The curves are fits to Eq.\ \ref{crit-fit} over the range $25<T<T_N-0.4$~K, with $\beta$ fixed to 3D Heisenberg value 0.37.  The fitted values for $T_N$ are shown vs thickness in the inset.
}
\label{fig:crit}
\end{figure}

\begin{figure}
        \includegraphics[width=3.2in]{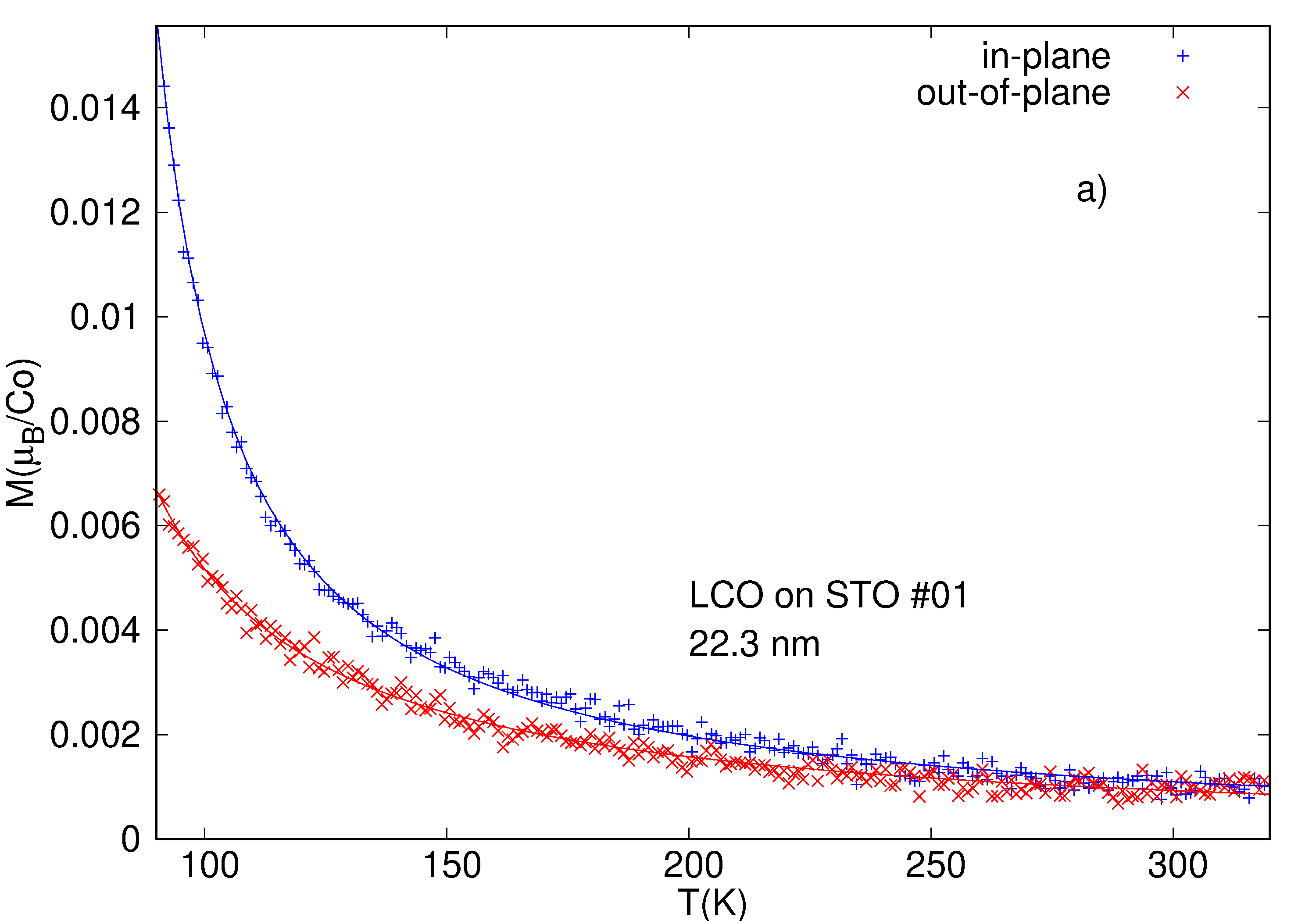}
        \includegraphics[width=3.2in,angle=0]{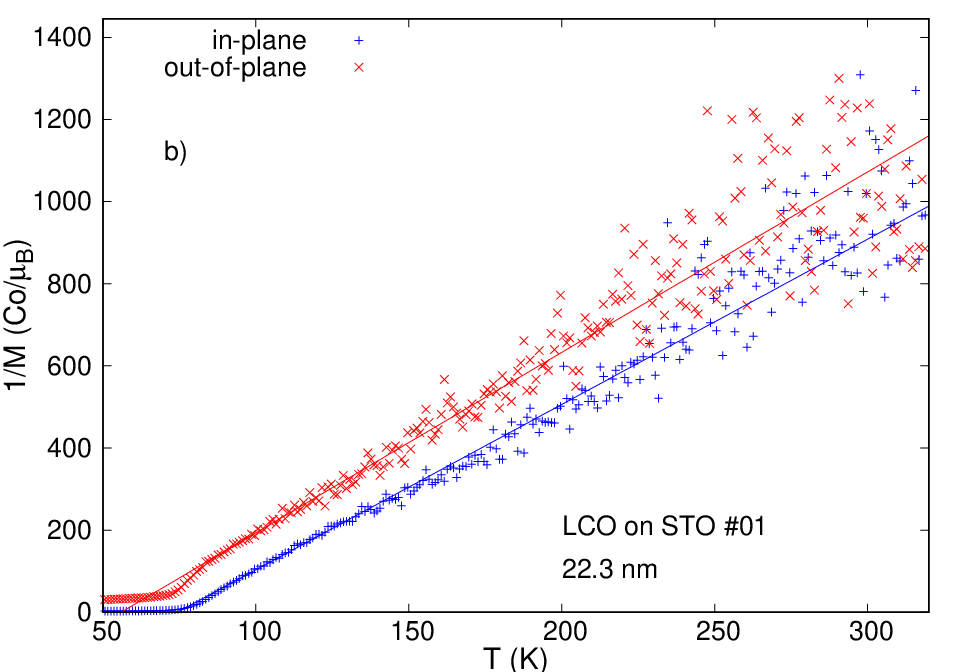}
        \caption
        {$M$ vs $T$ for a 22.3~nm thick LCO film on a STO substrate with $\mu_0 H=100$~mT.  Figure (a) shows the film and substrate data fit to Eq.\ \ref{CW_fit} after subtracting a fit of the substrate data to Eq.\ \ref{CW_fit}. Figure (b) shows $1/M$ vs $T$ to illustrate the straight line behavior and positive Curie-Weiss temperature.
}
\label{fig:22.3}
\end{figure}

\begin{figure}
        \includegraphics[width=3.2in]{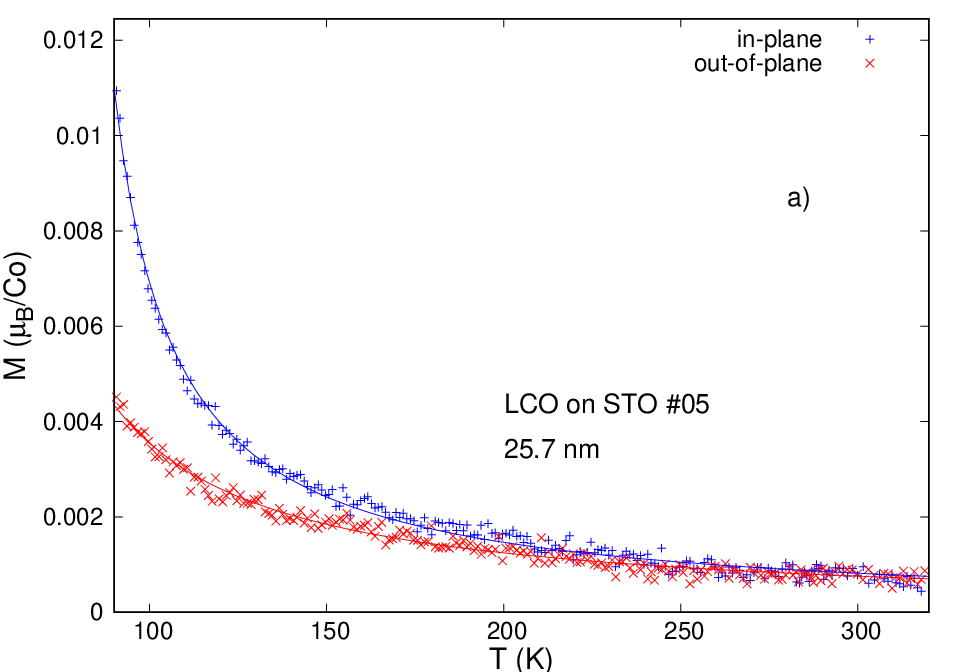}
        \includegraphics[width=3.2in]{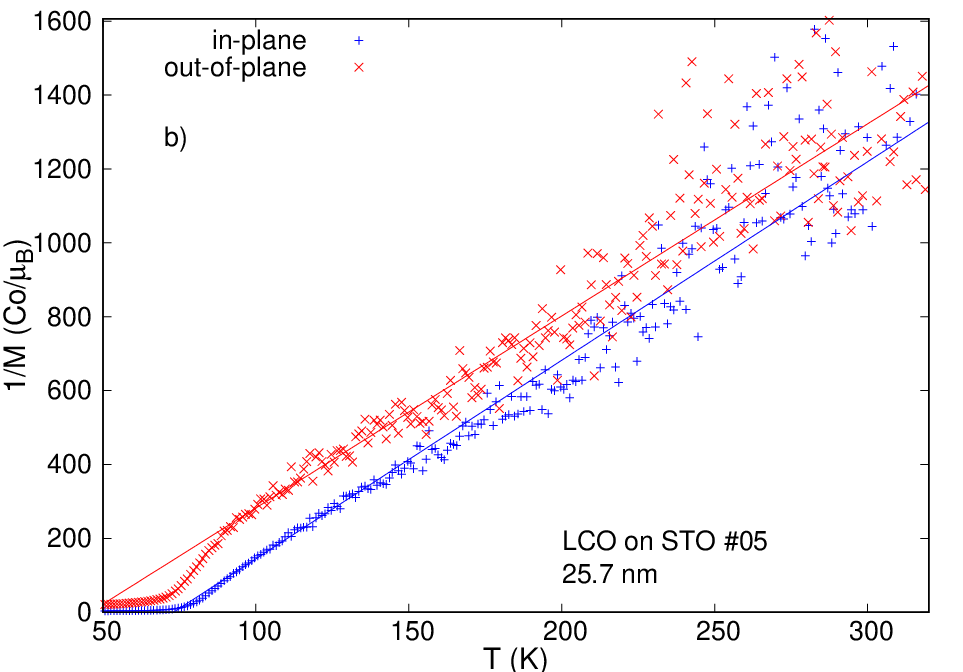}
        \caption
        {$M$ vs $T$ for a 25.7~nm thick LCO film on a STO substrate with $\mu_0 H=100$~mT.  Figure (a) shows the film and substrate data fit to Eq.\ \ref{CW_fit} after subtracting a fit of the substrate data to Eq.\ \ref{CW_fit}. Figure (b) shows $1/M$ vs $T$ to illustrate the straight line behavior and positive Curie-Weiss temperature.
}
\label{fig:25.7}
\end{figure}

\begin{table}

        \caption{LCO/STO fit parameters and moments using Eq.\ \ref{CW_fit} and Eq.\ \ref{crit-fit}.}
        
\begin{tabular}{|c|c|c|c|c|}
\hline

	sample            &  thickness (nm) & $C$ & $\theta_{CW}$ & $\mu _{\text{eff}}(\mu_{B})$   \\ \hline
	LCO01 - out-of-plane     &22.3 & 0.037(1) & 56(1) & 3.18(5)\\
	 - in-plane    & & 0.0399(5) & 74(1) & 3.33(2)\\ \hline
	LCO05 - out-of-plane     &25.7 & 0.041(2) & 46(2) & 3.37(8)\\
	 - in-plane    & & 0.0396(8) & 73(1) & 3.31(3)\\ \hline

\hline

\end{tabular}
\label{table:film_moment_analysis}
\end{table}

Figure \ref{fig:LCO10_LCO04} demonstrates the effect of reducing the film thickness.  For a film of 26~nm [Fig.\ \ref{fig:LCO10_LCO04}(a)], the spin-flop-like signal is still apparent, but the magnetic signal per Co is reduced relative to the 67.5~nm film.  With a further reduction to a thickness to 14.7~nm [Fig.\ \ref{fig:LCO10_LCO04}(b)], there is no evidence of spin-flop ordering.  Although the thermal stability was not as high, so the substrate signals are not as accurately subtracted as for the 67.5~nm, the decrease in ordering with thinner films is apparent.

The behavior of LCO films on STO substrates contrasts that of LCO films on LAO substrates.  Figure \ref{fig:crit} shows data from four LCO films with thicknesses 7.8, 10.3, 22.3 and 25.7~nm with the field in the plane of the film and perpendicular to it.  As discussed above, the significant nonmagnetic background must be subtracted over the $5<T<100$~K temperature range shown in the figure.  We first compensated for most of the background by subtracting the substrate data directly from the film plus substrate data.  Because this left a small background that varied from sample to sample, the data were shifted vertically to ensure they passed through zero at $T=100$~K, so the resulting plots mostly represent the magnetic ordering.  The measured magnetization perpendicular to the films is much smaller than in the plane and could simply be a result of imperfect alignment of the film with the film with the field or slight amount of film disorder.  A small sample to sample variation in the magnetization, on the other hand, is probably due to the shape anisotropy field of the thin film geometry as discussed in the Ref. \cite{Guoeaav5050}.

For the in-plane signal, $M$ vs $T$ is shown in Fig.\ \ref{fig:crit} along with fits to Eq.\ \ref{crit-fit}.  The order parameter critical exponent was fixed to the three-dimensional Heisenberg value 0.37 and the other parameters were allowed to vary.  The values of $T_N$ obtained from the fits are shown in the inset. The amplitudes vary somewhat. It has recently been shown that there can be effects at low temperature from crystal twins~\cite{gdkhwfl19}, but not near the transition temperature; the critical amplitudes might be more affected by slight disorder and initial cooling rates. The fits are reasonable, despite being done for temperatures $25<T<T_N-0.4$~K that include data likely well outside the asymptotic critical region. The compatibility with the three-dimensional Heisenberg critical exponent sharply contrasts the behavior found for bulk LCO, where $\beta \approx 0.7$ resulting from surface ordering at twin interfaces  \cite{bh72,bh74,dhbcyfab15_2,kbyfwmy18}.

For temperatures above $T=100$~K, the magnetic signal is too weak in the 10.3 and 7.8~nm films to allow reliable fits, but good fits were obtained in the 22.3 and 25.7~nm films.  Excellent fits to Eq.\ \ref{CW_fit} between $T=120$ and $320$~K were obtained for the two LCO films on STO after subtracting fits to the substrate data, as shown in Fig.\ \ref{fig:22.3} (a) and Fig.\ \ref{fig:25.7} (a).  The same results are shown as $1/M$ vs $T$ in Fig.\ \ref{fig:22.3} (b) and Fig.\ \ref{fig:25.7} (b) to illustrate the straight line behavior at higher temperatures and the consistent behavior of the two samples.  Values for the effective moment per Co ion, $\mu_\text{eff}$, are calculated using
\begin{equation}
    \mu_\text{eff} = \left(\frac{3k_{B}C}{\mu_{B}^2N_{A}}\right)^{1/2},
    \label{mu_eff}
\end{equation}

\noindent where $\mu_{B}$ is the Bohr magneton, $k_{B}$  Boltzmann's constant and $N_{A}$ is Avogadro’s number.  The parameters $C$ and $\mu_\text{eff}$ obtained from the fits, (Table~\ref{table:film_moment_analysis}) are fairly consistent for the two samples and the two field orientations, and the values for $\mu_\text{eff}$ are close to the value 3.45(2)$\mu_{B}$ obtained for bulk LCO \cite{dbbycfb13}. The positive values of $\theta_{CW}$ are consistent with the ordering of a net moment; they do not necessarily indicate ferromagnetic interactions. In ferromagnets, where the applied field is conjugate to the order, the Curie-Weiss parameter is positive and roughly reflects the net strength of the
interactions and, thus, the transition temperature. The applied field normally does not couple directly to antiferromagnetic ordering and, consequently, the Curie-Weiss temperature is negative with a magnitude roughly corresponding to the transition temperature.  However, when the magnetic moments in an atiferromagnet are canted as a result of a structural distortion, the field does couple directly to the ordering and the Curie-Weiss temperature will have a positive value with a magnitude reflecting the strength of the net antiferromagnetic interactions. Such is the case of a LCO thin film on a STO substrate.  Good fits to the magnetization well above $T_N$ are possible in this case because the positive values of $\theta_{CW}$ result in a significant variation with temperature in the fitting region.

\begin{figure}
        \includegraphics[width=5in]{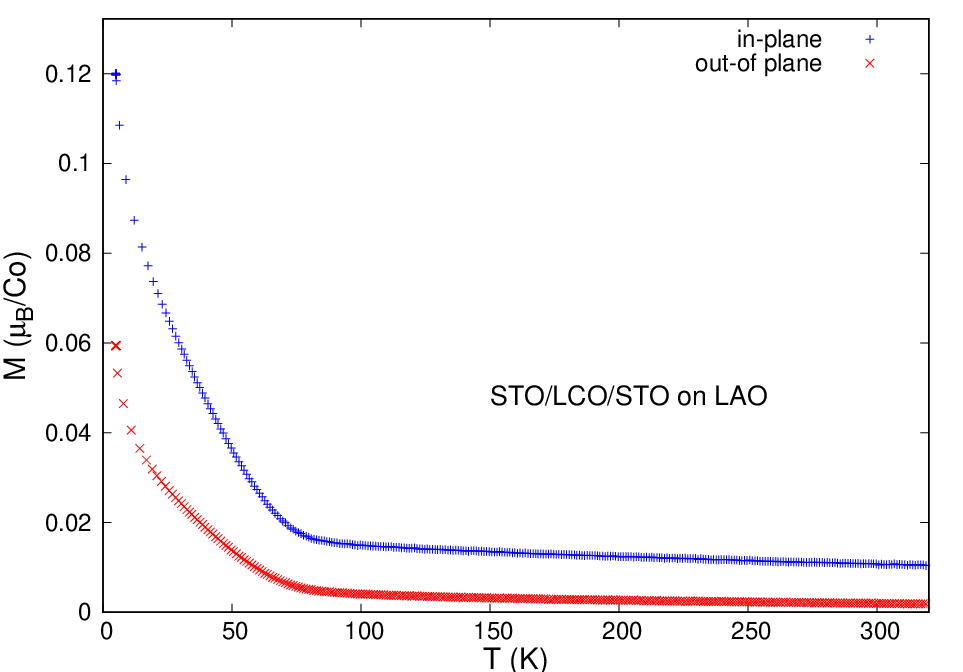}
        \caption
            {$M$ vs $T$ for a LCO/STO multilayer film with a top 13.9~nm STO layer, a 40.7~nm LCO layer, a 37.8~nm STO layer and a 9.4~nm LCO layer on a LAO substrate with $\mu_0 H=100$~mT.  The non-magnetic background was roughly subtracted by fitting the data for $T>150$~K to Eq.\ \ref{CW_fit}.  As we discussed in the text earlier, the LAO substrate has a complicated temperature dependence and this is further complicated by the STO films and the STO/LAO interactions. However, it is clear that antiferromagnetic ordering takes place without a net moment and is stronger than the LCO films on LAO substrates.
}
\label{fig:LCO20}
\end{figure}

To explore whether the interface chemistry plays a role in the nature of the ordering, and to see the effect of imposed lattice constraints intermediate to those imposed by the LAO and STO substrates, the magnetization vs $T$ of a multilayer consisting of a 13.9~nm STO capping layer, a 40~nm LCO layer, a 37.8~nm STO layer and a 9.4~nm LCO buffer layer grown on a LAO substrate was measured with an external field $\mu_0 H=100$~mT in the plane of the sample and perpendicular to it.  Although LCO interfaces directly with STO, but not LAO, there is no evidence of ordering involving a net moment.  The Bragg peak is broad with a half width of 1.2$^\circ$, (Fig. S5, supplemental Material) indicating a significant spread in $c$ parameter values, but a fit to the peak position yields an average $c$ value intermediate between the cases of LCO interfacing directly with a LAO and interfacing with STO, as shown in Table~\ref{table:film_parameters}.  It appears that the operating influence on the LCO film is not the chemistry of the interface, but rather the degree of strain distorting the LCO lattice.  This is consistent with the voltage-controlled strain used to switch the net moment on and off in a LCO layer of a multilayer device \cite{hppjdy13}.  

\subsection{The Correlation Between the Co-O-Co Bond Angle and the Magnetization}

Insight into the behavior of the LCO films can be gained by making comparisons to bulk LCO, where antiferromagnetic interactions are highly correlated with the Co-O-Co angle and are suppressed for angles below 163$^\circ$.  The experimental observations are consistent with calculations predicting the critical angle necessary for magnetic interactions \cite{lh13}.  Although that correlation has not been calculated for the film environment where LCO is distorted by sublattice strain, it appears from the observed behavior in films discussed below that the general magnetic behavior approximately conforms to the same dependence of the magnetic interactions on the Co-O-Co angle.

Although the lattice parameters of the films have been measured only at room temperature, we can estimate the structure at low temperature assuming similar relative changes in the $a$ and $c$ pseudocubic lattice parameters upon cooling as observed in bulk LCO.  LCO parameters shorten by approximately 1\%\ (as do the parameters in LAO and STO \cite{llufgtkss10,HMRSCKGSDZC05}).  The Co-O bond length in LCO shortens by about $0.5$\%\ upon cooling from room temperature to low temperature.  The most relevant effect of cooling from room temperature in bulk LCO is to reduce the Co-O-Co angle by about one degree, which is important because the bulk Co-O-Co angle at room temperature is near the critical value of 163$^\circ$ and it decreases to that critical value near $T=40$~K.

Neutron and x-ray scattering and EXAFS techniques have been used to study the details of the temperature dependence of the lattice parameters in bulk LCO \cite{bkbdmzmb16,jbsbamz09}.  EXAFS data were also used to determine the Co-O bond length in LCO films on LAO and STO substrates \cite{srkkmsw12} along directions parallel and perpendicular to the substrate interface.  When possible, lattice parameters are most accurately determined using scattering techniques, but the amount of material in thin films makes an accurate determination of the Co-O bond length difficult.  However, the EXAFS technique has been used to accurately determine the ratio of the Co-O bond lengths that are approximately along the $c$ and $a$ directions.   For LCO on LAO, the ratio of the Co-O bond length along $c$ to that along $a$ is determined to be 1.003(4), which is close to unity, and the ratio for LCO on STO is 0.979(6), which significantly deviates from unity.  Because the LCO films have the same chemical environment as bulk LCO, we assume in our discussion that the Co-O bond lengths in the films are only altered when necessary to conform to the lattice mismatch between the film and substrate.

We assume that the Co-O bond lengths for LCO films on LAO are close to the the bulk LCO value of 1.924(2)~\AA{} at room temperature and 1.915(2)~\AA{} at low temperature obtained from other studies \cite{sjabbbmpz09,dbbycfb13,dbhycfbab15}. However, the average measured $c$ lattice parameters for the film, listed in Table~\ref{table:film_parameters}, is 3.87~\AA, which is slightly larger than twice the room temperature bulk Co-O bond length of 3.85~\AA{}.  Note, however, that in bulk LCO, the $c$ lattice parameter decreases slightly more than the Co-O bond length as the temperature decreases, so the film $c$ parameter and the Co-O bond length possibly agree even better at low temperatures. For the $c$ lattice parameter to be twice the bulk Co-O bond length, or perhaps slightly stretched to fit the Co-Co bond length, suggests that the Co-O bond is close to being aligned with the $c$ direction and the Co-O-Co bond angle is close to 180$^\circ$.  If Co-O bond length is actually one or two percent larger than the $c$ parameter, the Co-O-Co angle will not be significantly smaller than 180$^\circ$; it will not be close to the critical angle of 163$^\circ$.  Assuming the other two Co-O bonds are perpendicular to this one, they must lie nearly in the plane parallel to the substrate interface.  To accommodate the room temperature lattice parameter $a=3.791$~\AA, the Co-O-Co bond angles must be close to 160$^\circ$ and we would expect this angle to decrease upon cooling as it does in bulk LCO.  In bulk LCO, Co-O-Co angles less than 163$^\circ$ prevent magnetic interactions \cite{bkbdmzmb16}.  If we assume a similar cutoff angle in the LCO films on LAO substrates, we would expect the interactions between Co sites along the direction perpendicular to the substrate interface to be strong, but there should be no significant magnetic interaction in the plane parallel to the interface.  This magnetic interaction geometry would support a one-dimensional magnetic system, but one-dimensional magnets cannot exhibit long-range order.  This is consistent with the observation from magnetometry experiments that, in the thinnest films, no phase transition to long-range order takes place. To form a Co-O-Co angle greater than 163$^\circ$ to support antiferromagnetic interactions, the Co-O bonds parallel to the substrate interface would need to be 1.917~\AA{} or smaller at room temperature.  There is no reason for Co-O bond to shorten from the room temperature bulk value, however, because that is not required to accommodate the substrate lattice parameter. 

In the thicker LCO films on LAO substrates, a spin-flop-like transition is observed and the moment at low temperature grows with increasing thickness.  The thicker films also exhibit a new peak in the Bragg scattering, as shown in Fig.\ \ref{fig:XRD2}, that suggests a relaxation of the lattice towards bulk LCO lattice parameters. Table \ref{table:film_parameters} shows that the secondary peak corresponds to a vertical $c$ parameter of~3.818~\AA, close to the bulk value.  Interestingly, the change in the lattice structure appears abrupt rather than continuous.  AFM images (Fig. S6, supplemental Material) show the surface of the film acquiring granular features and this is likely that results from relaxed non-epitaxial material with an altered structure, similar to the NbO$_2$ epitaxial films discussed elsewhere \cite{JCML19}.  Once the lattice parameters allow for Co-O-Co angles larger than about 163$^\circ$ in all directions, three-dimensional magnetic ordering is possible.  The behaviors of the thin and thick films prove consistent with the geometric model for the correlation of the magnetic behavior with the Co-O-Co critical angle in bulk and nanoparticle LCO.

LCO films on STO substrates represent a quite different strain configuration.  Instead of compressing the Co-O bonds along the $a$ direction, they are significantly stretched.  The room temperature $a$ lattice parameter is 3.905~\AA, significantly larger than 3.85~\AA, twice the bulk LCO Co-O bond length of 1.924(2)~\AA.  Transmission electron microscopy (TEM)  measurements \cite{lwzrsjcwwghggjg19} confirm that the Co-O-Co angle is close to 180$^\circ$ in that plane and that condition is likely unchanged upon cooling.  It is unlikely that the Co-O bond could be even larger than required to fit along the $a$ direction of the lattice.  If the third Co-O bond is perpendicular to the other Co-O bonds, it must be significantly compressed relative to the others and to the bulk Co-O bond length.  The $c$ lattice parameter for films LCO01 and LCO05 is 3.79~\AA, which is significantly smaller than twice the bulk LCO Co-O bond length, 3.85~\AA, by a ratio of 0.984. XAFS measurements \cite{srkkmsw12} indicate the ratio of the Co-O bond length along $c$ to that along $a$ is 0.979(6), which is consistent with the compression of the Co-O bond needed to fit the $c$ parameter and suggests that the Co-O bond is nearly aligned with the $c$ axis.  Upon cooling, the lattice structure might change slightly, but clearly all of the Co-O-Co angles are much larger than 163$^\circ$ if all the Co-O bonds are perpendicular to each other.

The thinner films on STO substrates, LCO02 (10.3~nm) and LCO06 (7.8~nm), show slightly larger compression in the $c$ direction, with a ratio of $c$ to $a$ equal to 0.96, than LCO01 (22.3nm) and LCO05 (25.7 nm), but the net moments at low temperature are comparable; all four samples show a net moment between 0.30 and 0.44 $\mu_B$.  In Fig.\ \ref{fig:XRD2} (c), a small peak is evident just below 76$^\circ$ for LCO01 (22.3 nm), approximately at the angles of the film peaks of thinner LCO02 (10.3 nm) and LCO06 (7.8 nm) films.  This might represent the layers adjacent to the substrate of the thicker film that have not relaxed as much.

Unequal Co-O bonds along the $a$ and $c$ directions should result in the loss of cubic symmetry experienced by the magnetic moments centered on the Co sites.  It is also possible that the perpendicular Co-O bonds are not exactly along the $c$ axis to avoid some of the compression.  That would also result in the loss of cubic symmetry.  In either case, with the loss of cubic symmetry, the magnetic moments associated with each Co site could misalign with the $c$ axis and, with antiferromagnetic ordering, the tilt direction would then alternate from layer to layer.  Twin domains with 90$^\circ$ rotations would also be expected with a tilt and twinning has been observed \cite{gdkhwfl19}.  The resulting non-collinear alignment of the moments would result in a net moment perpendicular to the $c$ axis, a condition consistent with the observation from magnetometry experiments that the net moment is predominantly, if not entirely, in the plane of the interface.  It is likely that the origin of the net moment in an LCO film on a STO substrate is the canting of the antiferromagnet sublattices.  If the magnetic sublattices create a net moment, we can estimate the angle of tilt from the $c$ axis by comparing net moment achieved at low $T$ with the moment on each Co site.  The net moment per Co site is 0.44~$\mu_B$, whereas the moment on each Co, from Table \ref{table:film_moment_analysis} is approximately 3.3~$\mu_B$, which corresponds to a tilt angle as large as 8$^\circ$, assuming moments at all Co sites contribute to the long-range order.

A resonant x-ray scattering study~\cite{snmskwpkr18} of a LCO film on a STO substrate was interpreted to suggest charge ordering at and below room temperature, where adjacent Co sites are in alternating low spin (LS) and high spin (HS) states, and the magnetic interactions are ferromagnetic and exist only between second-nearest HS neighbors.  The experimental data were interpreted in the context of a DFT calculation where such a spin-state ordering is imposed.  Although the calculations and resonant scattering results are internally consistent, the \textit{a priori} imposition of spin-state ordering does not necessarily exclude a model where the interactions are antiferromagnetic and similar to bulk LCO, but are associated with distorted oxygen octahedra.  It would be useful to compare DFT calculations under these conditions with the resonant x-ray scattering data \cite{snmskwpkr18}.

\section{Conclusions}
We have shown how the appearance in LCO films of a net moment resulting from noncolinear antiferromagnetic sublattices correlates well with the Co-O-Co angle. That correlation provides insight into why LCO films can exhibit a net moment or not depending on the strain from the substrate.  It also demonstrates that a detailed understanding of the magnetic interactions in bulk LCO is key to understanding them in thin films. In our experiment, LCO films grown on LAO substrates where LAO provided a compressive strain to the film ($a/c < 1$) had no magnetic ordering for strained films and spin-flop-like ordering for partially relaxed thicker films. The films grown on STO substrates with a biaxial tensile strain ($a/c > 1$), on the other hand, showed an abrupt increase in moment below 80 K. M vs T signals from the films grown on STO followed the critical behavior with the critical exponent of Heisenberg 3D value (0.37) indicating a 3D ordering in contrast to the surface ordering reported in the bulk LCO.  The effective moment per Co ion ($\mu_{eff}$), however, remained consistent with the bulk LCO.  

It has been demonstrated that substrate strain can be used to switch between states with or without net moments in a LCO film, for example by applying voltages to an adjacent SrTiO$_3$ film; this forms the basis of a low temperature logic device \cite{hppjdy13}.  Establishing that antiferromagnetism correlates well with the Co-O-Co angle opens up a new possible path to the development of room temperature switchable magnetic devices using oxide materials.  If an antiferromagnet film can be strained in a way that introduces misalignment of the sublattices, it can gain a significant net moment.  If that strain can be altered appropriately and the Co-O-Co angle is near a critical value, that net moment can be switched.  Such a strain-driven switch mechanism does not require introduction of ferromagnetic interactions into the system.

\section{Acknowledgments} 
 The work at UCSC was supported by the Air Force Office of Scientific Research under award number FA9550-19-1-0307 and the University of California, Multicampus Research Programs and Initiatives grant MRP-17-454963. Work at the SSRF was supported by National Key Research and Development Program of China (2017YFA0402800) and National Science Foundation of China (NSFC grants: U1732121, U1932201). We thank Frank Bridges for useful discussions and Bin Zhao, Xiaolong Li and Ronaldo Rodriguez for technical assistance with the experiments.

\bibliography{magnetism.bib}  
\
\
\section{Supplemental Material} \
This supplemental document includes following figures: \

Figure 9. S1: X-ray diffraction spectra of LaCoO$_3$ samples\

Figure 10. S2: Reciprocal space map of (103) reflection of a 25.7 nm LaCoO$_3$ film\

Figure 11. S3: Atomic force microscopy images of LaCoO$_3$/SrTiO$_3$ samples\

Figure 12. S4: X-ray reflectivity spectra of LaCoO$_3$ samples\

Figure 13. S5: XRD and XRR of the LaCoO$_3$/SrTiO$_3$ multilayer sample\

Figure 14. S6: AFM image of a relatively thicker LaCoO$_3$/LaAlO$_3$ film\

\begin{figure}
    \centering
    \includegraphics[width=5.1in]{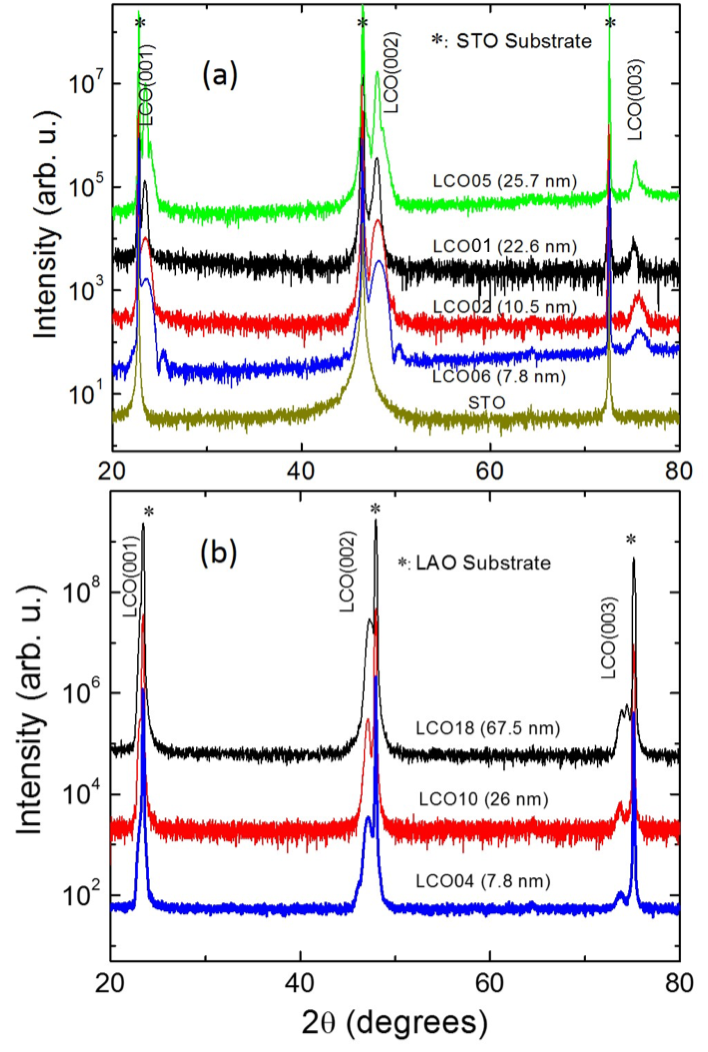}
    \caption {S1: X-ray diffraction spectra (XRD) of LCO films with different thicknesses grown on STO (a) and LAO (b) substrates with 2$\theta$ ranging from 20 to 80$^\circ$.}
    \label{*}
\end{figure}

\begin{figure}
    \centering
    \includegraphics[width=4in]{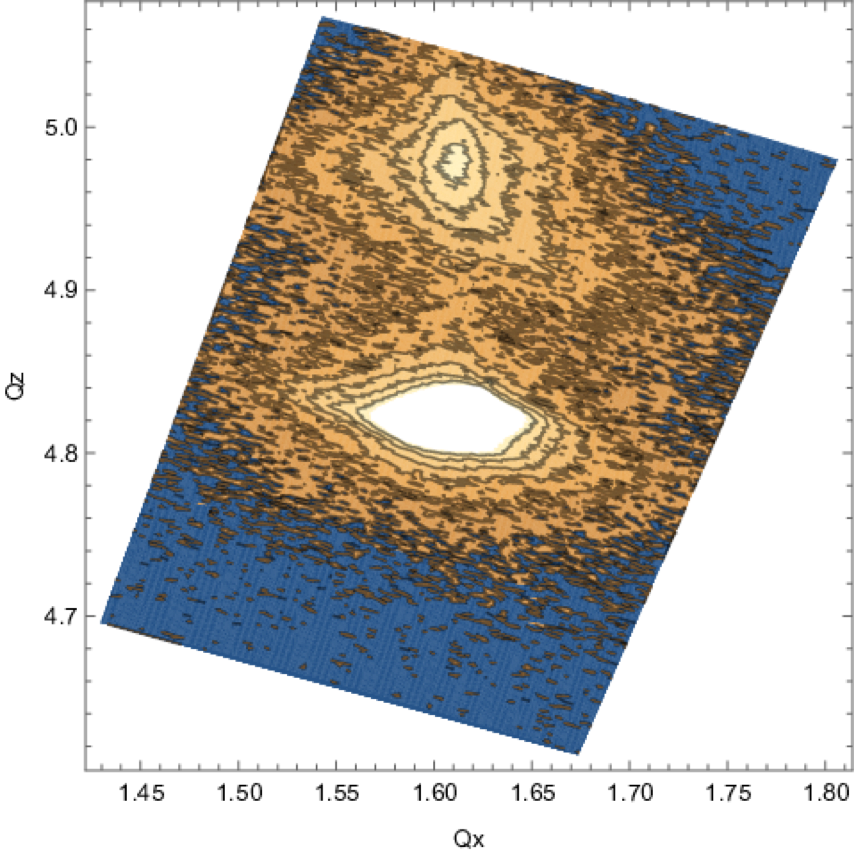}
    \caption        {S2: Reciprocal Space map of (103) reflection of LCO05 (25.7 nm) film.}
\label{*}
\end{figure}

\begin{figure}
\centering
        \includegraphics[width=4in]{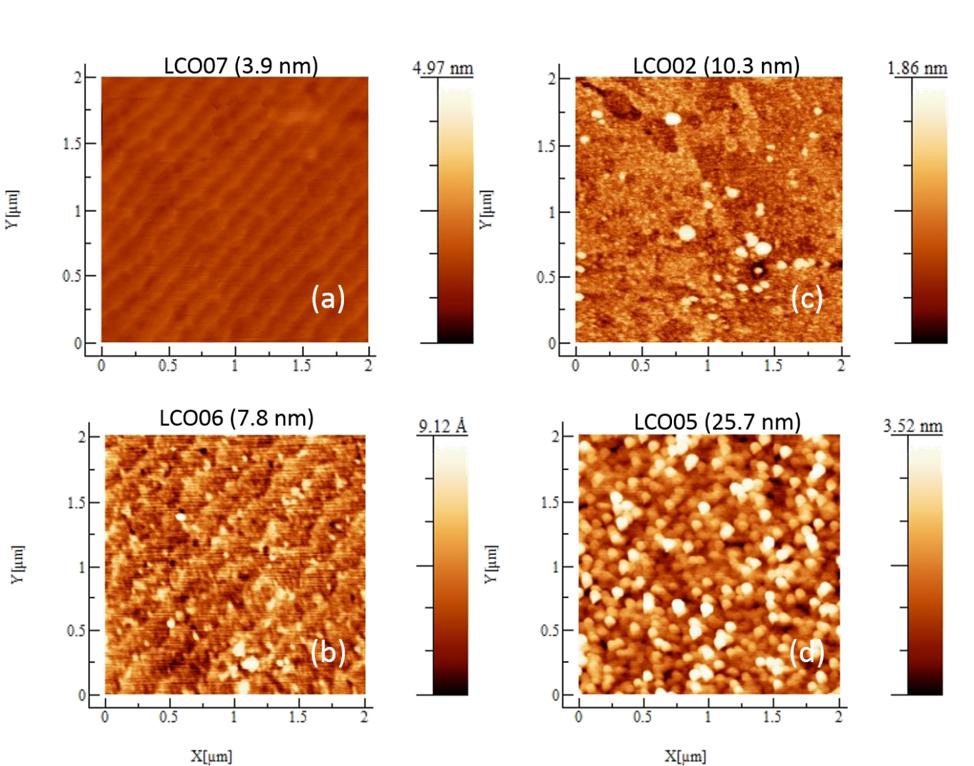}
        \caption
        {S3: AFM images of LCO samples grown on STO: (a) an ultrathin film with the atomic terraces from the STO substrates. AFM images of (b) 7.8 nm (c) 10.3 nm (d) 25.7 nm film. With increased film thickness, the film surface gets rougher and the step retraces become less obvious.
 }
 \end{figure}
 
\begin{figure}
\centering
        \includegraphics[width=7in]{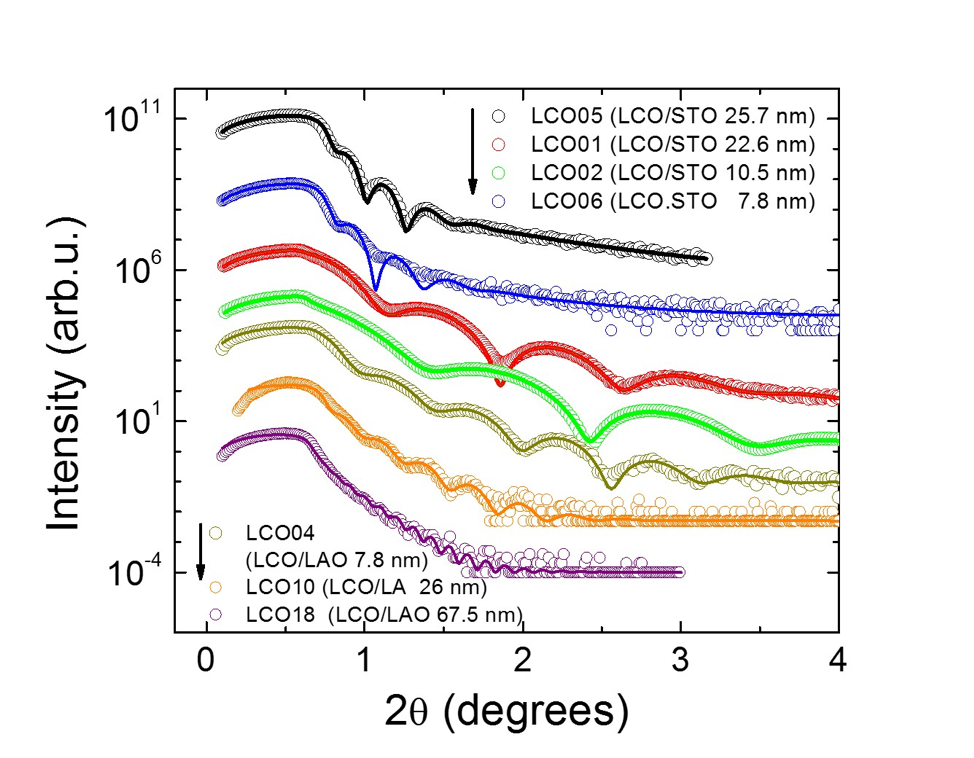}
        \caption
        {S4: X-ray reflectivity of the samples grown on STO and LAO substrates. The circular symbols are the data and solid curves are the simulations. The GenX software was used to fit and extract the thickness of the films. 
}
\end{figure}

\begin{figure}
\centering
        \includegraphics[width=5.1in]{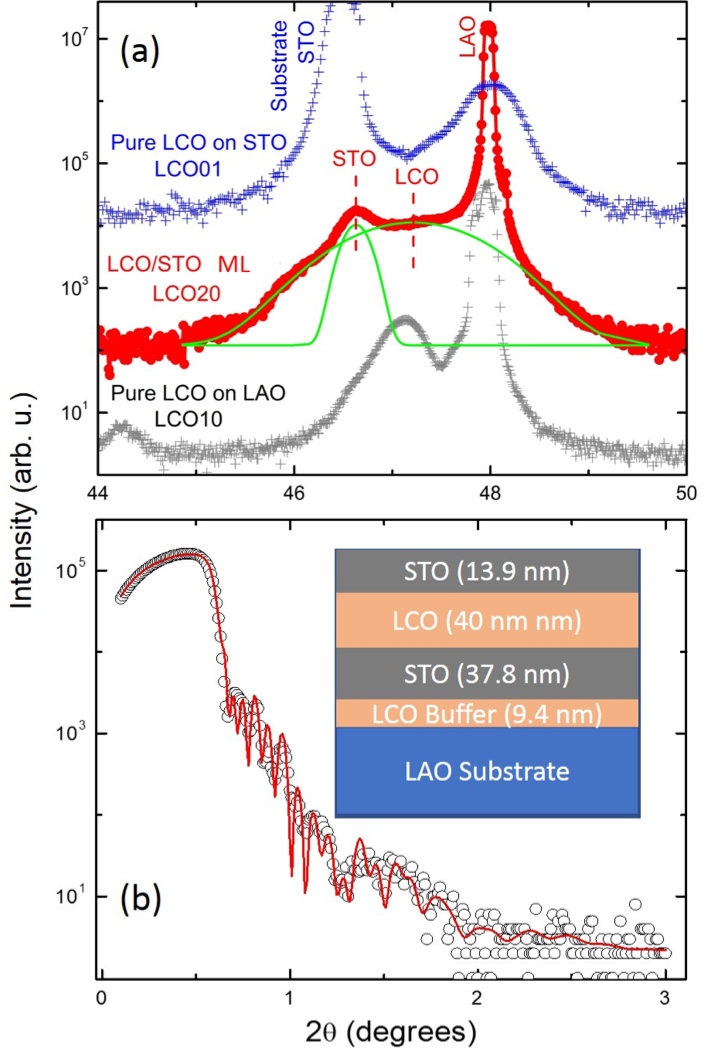}
        \caption
        {S5:  (a) X-ray diffraction from an LCO/STO multilayer [STO (13.9 nm) + LCO (40.7 nm) + STO (37.8 nm) + LCO (9.4 nm)]/STO sample. Two other films: LCO on STO (LCO01) and LCO on LAO (LCO18) are plotted together for comparison. Fig. (b) is the x-ray reflectivity with the simulation (solid curve). Inset shows the sample stacking structure. 
}
\end{figure}

\begin{figure}
\centering
        \includegraphics[width=5in]{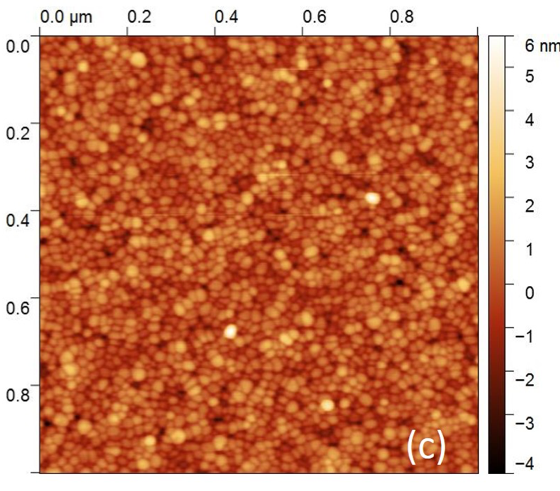}
        \caption
        {S6: (a) AFM image of the 67 nm thick LCO film on LAO substrate. With increased film thickness, the film surface becomes rougher with granular features. 
}
\end{figure}        
      
\end{document}